\documentclass[prb,reprint,superscriptaddress,amsmath]{revtex4-1}
\usepackage{graphicx}

\newcommand{\G}{ \Gamma }
\newcommand{\g}{ \gamma }
\renewcommand{\o}{ \omega }
\newcommand{\veps}{ \varepsilon }

\newcommand{\epspi}{ \varepsilon_{\mathbf{k}}^{\pi} }
\newcommand{\epspis}{ \varepsilon_{\mathbf{k}}^{\pi^*} }
\newcommand{\Deleps}{ \Delta\varepsilon_{\mathbf{k}} }
\newcommand{\bargam}{ \bar{\gamma}_{\mathbf{k}} }

\newcommand{\eF}{ \varepsilon_{\mathrm{F}} }
\newcommand{\eL}{ \omega_{\mathrm{L}} }
\newcommand{\eph}{ \omega_{\mathrm{ph}} }
\newcommand{\ecut}{ \varepsilon_{\mathrm{cut}} }

\newcommand{\oin}{ \omega_{\mathrm{in}} }
\newcommand{\oout}{ \omega_{\mathrm{out}} }
\newcommand{\oph}{ \omega_{\mathrm{ph}} }

\newcommand{\<}{\langle}
\renewcommand{\>}{\rangle}

\renewcommand{\k}{ \mathbf{k} }
\newcommand{\M}{ \mathcal{M} }
\newcommand{\Mk}{ \mathcal{M}_{\mathbf{k}} }

\newcommand{\mr}{ \mathrm }

\renewcommand{\AA}{\r A}

\begin{document}



\title{ {\it Ab initio} calculation of the $G$~peak intensity of graphene:
        Combined study of the laser and Fermi energy dependence and importance of quantum interference effects }


\author{Sven Reichardt}
\affiliation{Physics and Materials Science Research Unit, University of Luxembourg, 1511 Luxembourg, Luxembourg}
\affiliation{JARA-FIT and 2nd Institute of Physics, RWTH Aachen University, 52074 Aachen, Germany}

\author{Ludger Wirtz}
\affiliation{Physics and Materials Science Research Unit, University of Luxembourg, 1511 Luxembourg, Luxembourg}



\begin{abstract}

We present the results of a diagrammatic, fully {\it ab initio} calculation of the $G$~peak intensity of graphene.
The flexibility and generality of our approach enables us to go beyond the previous analytical calculations in the low-energy regime.
We study the laser and Fermi energy dependence of the $G$~peak intensity
and analyze the contributions from resonant and non-resonant electronic transitions.
In particular, we explicitly demonstrate the importance of quantum interference
and non-resonant states for the $G$~peak process.
Our method of analysis and computational concept is completely general
and can easily be applied to study other materials as well.

\end{abstract}

\maketitle



\section{Introduction}

Raman spectroscopy of graphene has been a subject of considerable interest
as it is a fast and non-destructive way for sample characterization.
The typical Raman spectrum of pristine graphene displays only two prominent peaks, the so-called $G$ and $2D$ peaks,
which are the result of a scattering process involving one and two phonons, respectively.
Despite this apparent simplicity, the Raman spectrum yields a large amount of information on,
amongst others, doping, strain, inter-layer interaction, and the underlying substrate~\cite{ferrari2007,malard2009,ferrari2013,reichardt2017}.
To understand the influence of these quantities on the Raman spectrum of graphene,
considerable effort has been made to understand the
shape~\cite{basko2008,venezuela2011,berciaud2013,hasdeo2014,neumann2016c},
width~\cite{graf2007,pisana2007,herziger2014,neumann2015b},
height~\cite{cancado2007,basko2009,basko2009b},
and position~\cite{maultzsch2004,ferrari2006,berciaud2008,pisana2007,mohiuddin2009,yoon2011}
of the $G$ and $2D$~peaks.

While a clear picture for the $2D$~peak has been established~\cite{thomsen2000,maultzsch2004b,venezuela2011,herziger2014},
a corresponding simple picture for the $G$~peak is still missing.
The latest theoretical effort by Basko~\cite{basko2009} focused on
an analytical tight-binding approach in the low-energy regime.
In particular, the question of which electronic transitions contribute to the $G$~peak
has still not been satisfactorily addressed.
This question is also intimately tied to the role of quantum interference effects in the $G$~peak process,
which has been probed experimentally by the tuning of destructive interference effects
via variation of the Fermi level~\cite{chen2011}.

To address these open questions, we present results of a fully {\it ab initio} calculation of the one-phonon Raman intensity of graphene.
Our general and flexible approach allows us to study the laser and Fermi energy dependence of the $G$~peak intensity
and enables us to analyze and understand the contributions from resonant and non-resonant electronic transitions.
In particular, we explicitly demonstrate the importance of quantum interference
and non-resonant states for the $G$~peak process in the low-energy regime.\\
\\
This paper is organized as follows:

In section~\ref{sec:theo}, we describe our approach to the calculation of one-phonon Raman intensities from first principles.
Subsection~\ref{ssec:comp} contains the computational details of our {\it ab initio} calculations,
while we give details on the diagrammatic approach to the calculation of the Raman matrix element in subsection~\ref{ssec:diag}.
This subsection also introduces the two main concepts needed for the subsequent discussion:
quantum interference effects and resonant contributions to the Raman matrix element.

The results of our calculations are presented in section~\ref{sec:res}.
In particular we discuss:
(A) the laser energy dependence of the $G$~peak intensity,
(B) its Fermi energy dependence for different laser energies,
and (C) which electronic transitions play a non-negligible role for the $G$~peak process.


\section{Theory}
\label{sec:theo}

In order to compute one-phonon Raman intensities from first principles, we use a perturbative approach organized via Feynman diagrams
to calculate the quantum mechanical amplitude, i.e., the scattering matrix element, for a one-phonon Raman process.
The quantities entering the expression for the matrix element, such as the electronic band structure,
the electron-light, and the electron-phonon coupling were obtained from {\it ab initio} calculations
on the level of density functional theory (DFT), as detailed below.

\subsection{Computational details}
\label{ssec:comp}

The electronic density in the ground state was calculated with the {\tt PWscf} code as included in the {\tt Quantum ESPRESSO} suite~\cite{giannozzi2009}
using an ultrasoft pseudopotential.
The exchange-correlation functional was approximated on the level of the generalized gradient approximation (GGA) in the
parametrization of Perdew, Burke, and Ernzerhof (PBE)~\cite{perdew1996}.
Integrals over the first Brillouin zone (BZ) were carried out on a regular mesh of 60$\times$60$\times$1 $\k$-points,
while the plane-wave cutoff was set to 80~Ry.
These values lead to converged results for a vacuum spacing of 14~\AA,
which separates periodic copies of the graphene sheet in the $z$-direction.
As graphene has semi-metallic character, occupations were smeared out using a 0.002~Ry Fermi-Dirac smearing.
For the lattice constant, we use a value of 2.46~\AA, obtained from structure relaxation.

The frequency of the doubly-degenerate optical phonon at $\G$ was fixed at the experimentally~\cite{lee2012}
obtained value of 1581.6~cm$^{-1}$ for pristine, freestanding graphene.
Since we are interested in the intensity of the one-phonon Raman peak,
we neglect the dependence of the phonon frequency on the Fermi level as this shift is only on the order of a few cm$^{-1}$ (see Ref.~\onlinecite{pisana2007}),
corresponding to a change of 1-2~meV.
This change is negligible compared to the dominating energy scale appearing in the expression for the Raman intensity set by the laser energy,
which is on the order of 1-4~eV for visible light.

The electronic band structure, the electron-light coupling matrix element in the dipole approximation,
and the electron-phonon coupling matrix elements were obtained on a coarse 12$\times$12$\times$1 $\k$-point mesh
and then interpolated to a finer 480$\times$480$\times$1 $\k$-point grid
with the help of maximally localized Wannier functions using modified forms of the
{\tt Wannier90}~\cite{mostofi2014} and {\tt EPW}\cite{giustino2007,ponce2016} codes.
We only consider the $\pi$ and $\pi^*$~bands of graphene in the interpolation as relevant optical transitions
to the $\sigma$-bands are forbidden by selection rules.

Obtaining the above mentioned quantities on a very fine $\k$-point grid is necessary
to obtain converged results for the Raman matrix element.
We checked the convergence of our calculations with respect to the fine $\k$-point mesh
by testing mesh sizes ranging from 12$\times$12$\times$1 to 960$\times$960$\times$1 $\k$-points.

\subsection{Diagrammatic calculation of Raman intensities}
\label{ssec:diag}

The intensity of the one-phonon Raman peak (the $G$~peak) per solid angle $\mr{d}\Omega$
can be calculated via a generalization of Fermi's golden rule~\cite{basko2009}:
\begin{equation}
\frac{\mr{d}I_G}{\mr{d}\Omega} \propto \frac{\oout^2}{(2\pi)^2c^4} \, |\M|^2 \times \delta(\oin - \oout -\oph),
\end{equation}
where
\begin{equation}
i\M 2\pi\delta(\oin-\oout-\oph) \equiv \< \oout,\nu; \oph,\lambda | i\hat{T} | \oin, \mu \>
\end{equation}
is a short-hand notation for the scattering matrix element.
We focus on the case of Stokes scattering and factored out the energy-conserving $\delta$-function from the matrix element.
For a phonon with finite lifetime, the $\delta$-function is to be replaced by a Lorentzian with a full width at half maximum of $\g_{\mr{ph}}$.
In the expression above, the state $| \oin, \mu \>$ represents an incoming photon with frequency $\oin$ and polarization $\mu$,
while the state $| \oout, \nu; \oph, \lambda \> \equiv  | \oout, \nu \rangle \otimes | \oph, \lambda \rangle$ is comprised of
one outgoing photon of frequency $\oout$ and polarization $\nu$ and one outgoing (in-plane) optical phonon with frequency $\oph$ and polarization $\lambda$.
Since the photon momentum is negligible, we use its frequency rather than its momentum to label the state. 
All photons are assumed to have momentum along the $z$-direction, i.e., to be polarized in-plane,
which is reasonable as the photon detector in an experiment is positioned on the axis of the incoming light perpendicular to the sample.

The operator $i\hat{T} = \hat{S} - 1$ denotes the non-trivial part of the scattering matrix ($S$-matrix).
Within the framework of diagrammatic, time-dependent perturbation theory, the scattering matrix element is given by~\cite{peskin1995}
\begin{widetext}
\begin{equation}
i\M \times 2\pi \delta(\oin-\oout-\oph) = \left[ \lim_{t\to(\infty-i\veps)} \< \oout,\nu; \oph, \lambda | 
         \mathcal{T} \exp\left( -i \int_{-t}^{+t}\mr{d}t'\, \hat{H}_{\mr{int}}(t') \right) | \oin, \mu \>
  \right]_{\substack{\text{connected,} \\ \text{amputated} \\ \text{diagrams only}} },
\end{equation}
\end{widetext}
where the subscript text indicates that, in an expansion of the exponential in terms of Feynman diagrams,
only the fully connected diagrams are to be retained and the propagators for the three external lines
(one each for the in- and outgoing photon and one for the phonon) are to be removed and replaced by the corresponding polarization vectors.
The operator $\hat{H}_{\mr{int}}$ in the formula above represents the sum of the interaction parts of the Hamiltonian (technically taken to be in the interaction picture),
i.e., it is the sum of the electron-photon and electron-phonon interaction Hamiltonian and $\mathcal{T}$ denotes the time-ordering symbol.

\begin{figure}[h!]
    \includegraphics{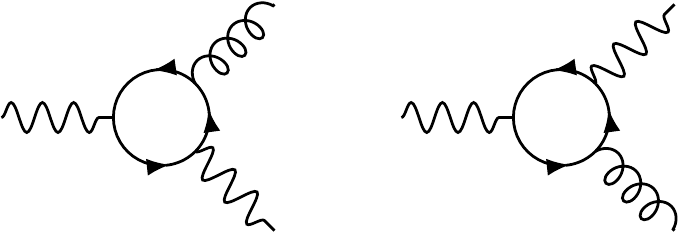}
    \caption{
        Leading order contributions in a diagrammatic expansion of the Raman scattering matrix element.
        Wavy lines represent photons, curly (gluon-type) lines phonons, and straight lines electrons.
    }
    \label{fig:diagrams}
\end{figure}

The leading order contributions in a diagrammatic expansion of the $T$-matrix element are shown in Fig.~\ref{fig:diagrams}.
These diagrams represent the leading order terms in the independent particle picture, i.e.,
correlation effects beyond those included in the one-particle Green's functions are neglected.
In particular, we neglect excitonic effects and correlation contributions to the electron-phonon coupling besides those included already on the DFT level.
For optical transitions in the visible regime, excitonic effects are assumed to be negligible in graphene.
Electronic correlation effects were shown to give a large contribution to the electron-phonon coupling for phonons with momenta near $\G$ and $K$~\cite{lazzeri2008}.
Since here we are only interested in processes involving the phonons at $\G$,
these corrections can, in a first approximation, be taken into account by a simple rescaling of the electron-phonon coupling~\cite{lazzeri2008}.
However, since  we will always compare the calculated Raman intensities or matrix elements to a calculated reference value (e.g., to an intensity at a certain $\oin$),
this correction factor will always cancel in the ratio and is thus not included here.

Explicitly, the contributions from the leading order Feynman diagrams read
\begin{widetext}
\begin{equation}
\begin{split}
i\M_1 &= \sum_\text{spin} \sum_{\k} \int \frac{\mr{d}\o}{2 \pi i} \mr{e}^{i \o 0^+} 
         \mr{tr}\left[ G_{\k}(\o) \g_{\k}^{\mu} G_{\k}(\o-\oin) \left( g_{\k}^{\lambda} \right)^{\dagger}
                       G_{\k}(\o-\oout) \left( \g_{\k}^{\nu} \right)^{\dagger} \right] \\
i\M_2 &= \sum_\text{spin} \sum_{\k} \int \frac{\mr{d}\o}{2 \pi i} \mr{e}^{i \o 0^+} 
         \mr{tr}\left[ G_{\k}(\o) \left( \g_{\k}^{\nu} \right)^{\dagger} G_{\k}(\o+\oout)
                       \left( g_{\k}^{\lambda} \right)^{\dagger} G_{\k}(\o+\oin) \g_{\k}^{\mu} \right].
\end{split}
\end{equation}
\end{widetext}
Here, $\g_{\k}^{\mu}$ and $g_{\k}^{\mu}$ are 2$\times$2 matrices in electronic band space whose entries
are the electron-light and electron-phonon coupling matrix elements between two electronic states at $\k$
for a photon/phonon with polarization $\mu$.
The matrix form of the electronic Green's function is denoted by $G_{\k}(\omega)$,
while ``$\mr{tr}$'' denotes the trace in the space of electronic bands.
Since all coupling matrices and Green's functions are diagonal in spin, the spin sum just yields a factor of 2.
The sum over $\k$-points is understood to represent an integration over the entire first BZ,
which we carry out by sampling the BZ with a regular mesh.
Satisfactory convergence was achieved with a 480$\times$480$\times$1-mesh.
Due to the fact that the total matrix element is a sum over $\k$-points,
contributions from different parts of the BZ can interfere constructively or destructively.
We will revisit this important point in section~\ref{sec:res}, when we discuss the interpretation of our results.

The frequency integral can be evaluated via contour integration, where the factor $0^+$
prescribes the closure of the contour in the upper half of the complex plane.
The application of the residue theorem leads to a sum of three expressions for each Feynman diagram,
i.e., six terms in total, which correspond to the six terms obtained from an expansion
in terms of time-ordered Goldstone diagrams
(compare, e.g., Ref.~\onlinecite{yu2010} for explicit expressions of the six terms).
After the frequency integration has been carried out,
the contributions to the matrix element at a specific $\k$-point can conveniently be grouped into three classes:
\begin{equation}
i\Mk = i\Mk^{\mr{aDR}} + i\Mk^{\mr{SR}} + i\Mk^{\mr{NR}},
\label{eq:Mres}
\end{equation}
where the superscripts ``aDR'', ``SR'', and ``NR'' label the almost double-resonant, the single-resonant,
and the non-resonant contribution to $\Mk$, respectively.
Here, ``resonant'' refers to resonance of an electronic transition with the frequency of the ingoing or outgoing light.
For instance, the aDR-contribution reads (in the case of zero doping):
\begin{widetext}
\begin{equation}
i\Mk^{\mr{aDR}} = \frac{ \Big( \left( \g^{\nu}_{\k} \right)^{\dagger} \Big)_{\pi,\pi^*}
                         \Big( \left( g^{\lambda}_{\k} \right)^{\dagger} \Big)_{\pi^*,\pi^*}
                         \Big( \g^{\mu}_{\k} \Big)_{\pi^*,\pi} }
                       { \left[ \oin - \Deleps^{\pi^*,\pi} + i\bargam^{\pi^*,\pi} \right]
                         \left[ \oout - \Deleps^{\pi^*,\pi} + i\bargam^{\pi^*,\pi} \right] }
\end{equation}
\end{widetext}
This amplitude describes a process in which an electron is excited from the $\pi$- to the $\pi^*$-band
and is subsequently scattered to an intermediate state in the $\pi^*$-band accompanied by emission of a phonon
before it finally radiatively recombines with the hole it left behind.
This contribution is maximal when the energy of an electronic transition
$\Deleps^{\pi^*,\pi} \equiv \epspis-\epspi$ equals the incoming or outgoing light frequency.
In particular, for laser energies $\oin \gg \oph$, we have $\oin \approx \oout$ and the aDR-term becomes quasi-double-resonant.
By contrast, the SR-term
\begin{widetext}
\begin{equation}
i\Mk^{\mr{SR}} = \sum_{s=\pi,\pi^*}
                    \left\{  \frac{ \Big( \left( \g^{\nu}_{\k} \right)^\dagger \Big)_{\pi,\pi^*}
                                    \Big( \g^{\mu}_{\k} \Big)_{\pi^*,s}
                                    \Big( \left( g^{\lambda}_{\k} \right)^\dagger \Big)_{s,\pi} }
                                  { \left[ \oout - \Deleps^{\pi^*,\pi} + i\bargam^{\pi^*,\pi} \right]
                                    \left[ -\oph - \Deleps^{s,\pi} + i\bargam^{s,\pi} \right] }
                           + \frac{ \Big( \left( g^{\lambda}_{\k} \right)^\dagger \Big)_{\pi,s}
                                    \Big( \left( \g^{\nu}_{\k} \right)^{\dagger} \Big)_{s,\pi^*}
                                    \Big( \g^{\mu}_{\k} \Big)_{\pi^*,\pi} }
                                  { \left[ \oph - \Deleps^{s,\pi} + i\bargam^{s,\pi} \right]
                                    \left[ \oin - \Deleps^{\pi^*,\pi} + i\bargam^{\pi^*,\pi} \right] } \right\}
\end{equation}
\end{widetext}
can only become single-resonant as the second factor in the denominator involves the phonon frequency.
In the expressions above, the quantity $\bargam^{s,s'}$ is the average of the decay widths of the states $|s,\k\>$ and $|s',\k\>$.
For simplicity, we use a constant value of 100~meV for the decay width of one state.

All other terms appearing in $i\Mk$ are summarily included in $i\Mk^{\mr{NR}}$.
These include terms that show resonant behavior with respect to the phonon frequency.
However, the set of electronic transitions form an area in $\k$-space that is much smaller than
the corresponding area for the transitions that are in resonance with respect to the incoming or outgoing light and its weight
in the sum over all $\k$-points is negligible and hence we do not include these terms in the single-resonant category.
We also want to point out that, at least in the case of graphene, the sum of the non-resonant contributions is by no means negligible,
as demonstrated in the next section.


\section{Results and discussion}
\label{sec:res}

The calculation of the Raman intensity using Feynman diagrams allows us to investigate the origin of various experimentally established facts about the $G$~line intensity.
We will first confirm that our model correctly predicts the strong dependence of $I_G$ on the laser frequency $\eL$ and show that this strong dependence
cannot be explained by the energy dependence of the joint density of states (JDOS) alone.

Secondly, we focus on the combined Fermi energy and $\eL$ dependence of $I_G$.
As measured and conceptually explained in Ref.~\onlinecite{chen2011}, the intensity of the $G$~line is greatly enhanced when $2\eF$ approaches a value of $\eL-\eph/2$.
Our first principles approach allows us to investigate this behavior as a function of $\eL$ as well,
going beyond the low-energy approximation used in the analytic work by Basko~\cite{basko2009}.
We find that the relative increase of $I_G$ at the ``critical'' values of $\eF$ depends non-monotonically on the laser energy.

Finally, we demonstrate that the common textbook approximation~\cite{yu2010} of retaining only the almost double-resonant term
in the Goldstone diagram expansion of the Raman amplitude fails in graphene.
Instead, at low laser energies, the full set of leading order Goldstone diagrams must be included to get quantitative agreement with the full result.
At higher energies, however, it is sufficient to approximate the full Raman amplitude with the sum of the almost double-resonant and the single-resonant terms.
Moreover, we demonstrate that the simple picture of only taking into account the resonant states (within a band of $\pm$ the electronic decay width), fails.
We find that the amount of states which contribute  to the total $G$~peak intensity depends strongly on the laser energy.
For low $\eL$, we can confirm the prediction of the analytical work of Basko~\cite{basko2009} that electronic states from almost the entire BZ contribute.
For larger values of $\eL$ ($\eL$$\gtrsim$2.5~eV), however,
we find that one needs to consider only a broad energy band of width $\approx$2~eV around the resonance energy window $[\eL-\eph,\eL]$
to obtain quantitative agreement with the full Raman amplitude.
By contrast, considering only the resonant energy band of width $\pm$100~meV
severely overestimates the matrix element as destructive quantum interference effects are neglected.

\subsection{Laser energy dependence}
\label{ssec:eL}

\begin{figure}[htb]
    \includegraphics{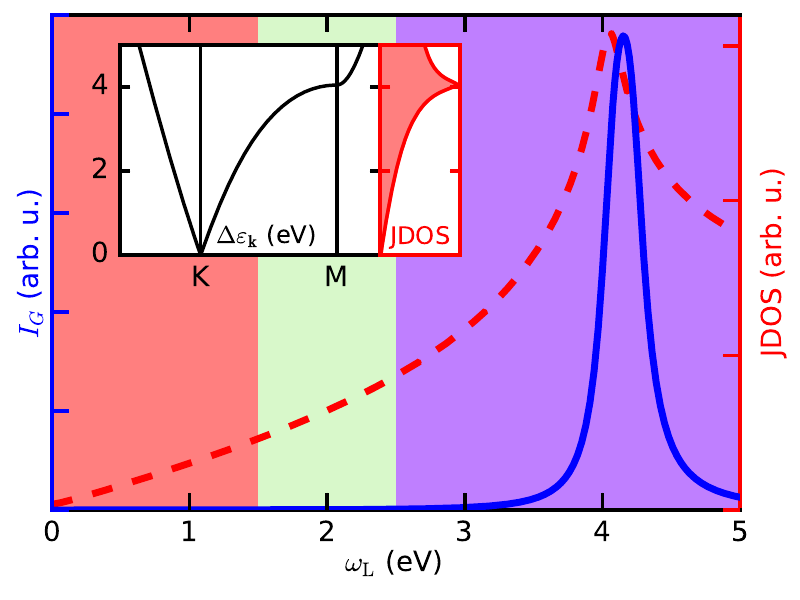}
    \caption{
        Intensity $I_G$ (blue line) vs. laser energy $\eL$.
        The red, dashed line represents the joint density of states (JDOS) as a function of transition energy.
        The shaded regions correspond to the three different regimes discussed in the text.
        Inset: Transition energy $\Deleps$ on part of the high-symmetry line $\G$-$K$-$M$-$\G$.
        The right panel shows the JDOS.
    }
    \label{fig:eL}
\end{figure}
The calculated laser energy dependence of $I_G$ is shown in Fig.~\ref{fig:eL}a (blue line).
We find a strong dependence of $I_G$ on $\eL$.
This behavior cannot be simply understood in terms of the joint density of states (JDOS) (red line).
The Raman intensity strongly peaks at a laser energy corresponding to the van Hove singularity of the JDOS at around 4.1~eV,
connected to the flatness of the transition energy band structure (see inset of Fig.~\ref{fig:eL}), as generally expected for optical spectra.
At low excitation energies, however, the intensity is suppressed even though a sizable number of bright electronic transitions is available.
This is in contrast to a simple absorption spectrum, which shows finite intensity whenever dipole-allowed electronic transitions are available.
To understand this more complex behavior of $I_G$ as a function of laser energy, we divide the excitation energy range of interest into three different regimes,
as indicated by the three colored shades in Fig.~\ref{fig:eL}a.

In the first regime, corresponding to laser energies $\eL$$\lesssim$1.5~eV, the Raman intensity is suppressed because of angular momentum conservation.
In the low-energy limit, the band structure appears to be circularly symmetric (``Dirac cone'' around the $K$~point, see inset of Fig.~\ref{fig:eL}).
Due to the continuous in-plane rotation symmetry in this regime, the $z$-component of the angular momentum is conserved
and thus the initial and final states must carry the same total angular momentum.
Since the electronic system is in its ground state both before and after the scattering event, its contribution to the total angular momentum can be ignored.
A (circularly polarized) photon carries angular momentum of $\pm\hbar$ and so does the created (circularly polarized) $E_{2g}$ phonon,
as it transforms like a two-dimensional vector under rotations.
Therefore the final state, consisting of one photon and one phonon can possess total angular momentum $+2\hbar$, $0\hbar$, or $-2\hbar$.
This differs from the angular momentum of the initial state (one photon) of $\pm\hbar$
and hence this process is disallowed by angular momentum conservation in the low-energy regime.

If one increases the laser energy to the second regime (1.5$\lesssim$$\eL$$\lesssim$2.5~eV), the band structure loses its circularly symmetric shape
and becomes trigonally warped.
The continuous rotation symmetry is broken down to the 120$^{\circ}$-rotation symmetry of the lattice
and exact angular momentum conservation is consequently reduced to angular momentum conservation up to integer multiples of $3\hbar$ only.
Thus an initial state with angular momentum $\pm\hbar$ can be scattered to a final state with total angular momentum $\mp2\hbar$.
As a consequence, incoming light with polarization $\sigma^{\pm}$ has opposite polarization after undergoing one-phonon Raman scattering.
The selection rules for linearly polarized light can be obtained from those for circularly polarized light by taking appropriate linear combinations
and can be conveniently expressed in terms of Cartesian Raman tensors~\cite{loudon1964}:
\begin{equation}
\mathcal{R}(x) =
    \begin{pmatrix}
        0 & c \\
        c & 0
    \end{pmatrix}, \quad
\mathcal{R}(y) =
    \begin{pmatrix}
        c & 0 \\
        0 & -c
    \end{pmatrix},
\end{equation}
where the argument in parenthesis refers to the polarization of the $E_{2g}$ phonon created
and the rows (columns) refer to the $x$- or $y$- components of the incoming (outgoing) light polarization.
We verified that our calculations respect these selections rules up to a relative factor of the order of $10^{-6}$.

\begin{figure*}[htb]
    \includegraphics{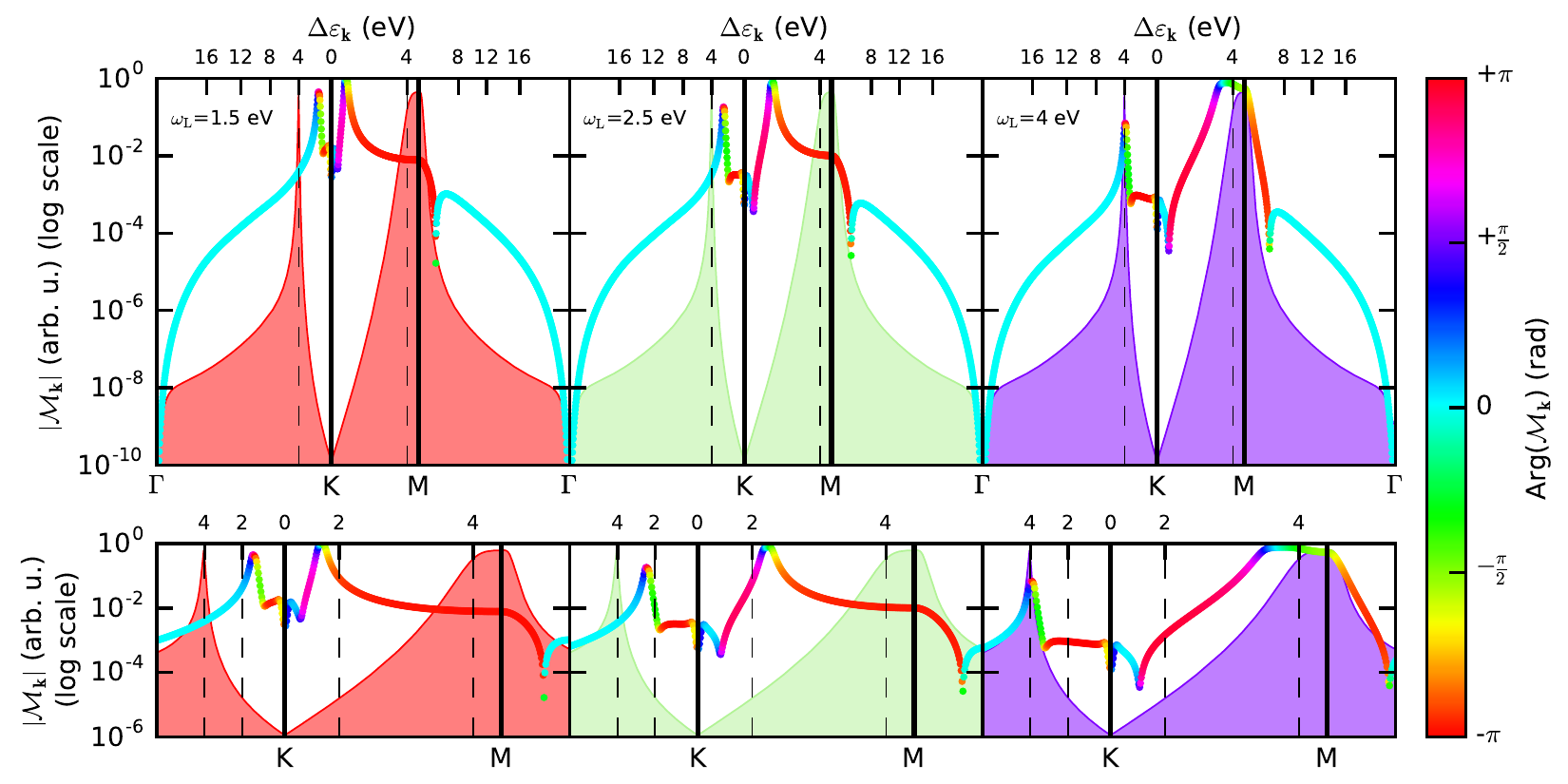}
    \caption{
        Absolute value (log scale) and phase (color-encoded) of the Raman matrix element $\Mk$ on the high-symmetry line $\G$-$K$-$M$-$\G$
        for $\eL$ = 1.5, 2.5, and 4~eV (left to right).
        The top horizontal axis displays the transition energy $\Deleps$ corresponding to the $\k$-point on the high-symmetry line.
        The shaded area represents the value of the JDOS at the transition energy $\Deleps$ of the corresponding $\k$-point.
        Lower panels: Zoom-in into the region between $K$ and $M$.
    }
    \label{fig:Mk}
\end{figure*}

To understand that the Raman intensity is still relatively small in the second regime,
we focus on the contribution of the different $\k$-points to the total Raman matrix element $\Mk = \sum_{\k} \Mk$.
To this end we look at the absolute value and phase of $\Mk$ on the high-symmetry line $\G$-$K$-$M$-$\G$ for three different values of $\eL$, as shown in Fig.~\ref{fig:Mk}.
The absolute value of $\Mk$ is given on the vertical axis and its relative phase is color-encoded.
The top horizontal axis displays the corresponding electronic transition energy $\Deleps$,
while the shaded background shows the value of the JDOS at this transition energy.
The latter is a measure for the weight of the corresponding $\k$~point in the total matrix element
when a full 2D~integration over the BZ is performed.

One can clearly identify the resonant states from the figure,
which correspond to the largest values of $|\Mk|$ (compare zoom-in shown in lower panels).
For both $\eL$=1.5 and 2.5~eV, the resonant states are sharply centered around one point on each side of $K$.
By looking at the phase of the resonant $\k$-points,
one notices that for each ``resonance peak'' the phase continuously undergoes a change of $\pi$ when passing over the peak,
i.e., contributions from $\k$-points to both sides of the resonant $\k$-point have opposite phase and cancel each other
(see blue to yellow dots in the first two panels of Fig.~\ref{fig:Mk}).
This change of phase by $\pi$ is typical for a driven system and is well-known from the simple system of a driven and damped harmonic oscillator.
It is of particular importance for the case of Raman scattering however,
as the total Raman amplitude is a sum over contributions from different $\k$-points and a relative phase of $\pi$ leads to destructive interference,
as seen for the states centered around the resonant $\k$-point.
While the contributions from the vicinity of the resonant $\k$-points to both sides of $K$ cancel separately,
it should also be noted that the ``inner'' flanks of the two resonance peaks in the $\G$-$K$ and $K$-$M$~directions
have opposite phase and also mostly cancel each other as they have similar amplitude and weight.
This is related to the residual continuous rotation symmetry.
A final point to note is the contribution of the non-resonant states near the van~Hove singularity in the vicinity of the $M$~point.
While the amplitude of these contributions is two orders of magnitude smaller than those of the resonant $\k$-points,
the corresponding $\k$-points are broadly spread along the $K$-$M$~direction of the high-symmetry line,
i.e., there are a lot of states along the high-symmetry lines with similar amplitude, which are also all in phase
(red dots around $M$ in the first two panels of Fig.~\ref{fig:Mk}).
Furthermore, when performing a 2D-integration over the entire first BZ,
these states enter with a lot of weight as the JDOS peaks at the corresponding transition energies.
Thus, the total contribution of the region around the van~Hove singularity at $M$, while non-resonant, is still far from negligible.
However, it turns out that this contribution is mostly canceled by the sum of the non-resonant contributions from the rest of the BZ
(cyan dots in the $\G$-$K$ an $M$-$\G$ directions in the first two panels of Fig.~\ref{fig:Mk}), which have opposite phase.
This fact will be demonstrated also (and more conclusively) in the last section of this paper.

This picture of dominating destructive quantum interference effects changes significantly for higher laser energy
($\eL$$\gtrsim$2.5~eV, see third shaded region in Fig.~\ref{fig:eL}a).
As seen in the third panel of Fig.~\ref{fig:Mk}, the contributions of the flanks of the two resonance peaks still destructively interfere with each other
(blue to yellow dots).
However, the resonance peak in the $K$-$M$~direction becomes very broad along the high-symmetry line and the contributions are mostly in-phase (cyan to green dots).
Combined with the high JDOS at the corresponding electronic transition energies,
the contributions from these resonant states dominate the total Raman matrix element $\M$ for higher laser energies.
This dominance is ultimately a consequence of the flatness of the transition band structure near the van~Hove singularity at the $M$~point (see inset of Fig.~\ref{fig:eL}).
Compared to the lower laser energy regime, here, the contributions from around the van~Hove singularity cannot be canceled 
by the sum of the non-resonant contributions from the bulk of the BZ, as the relative weight and amplitude of the resonant states
is much too large compared to those of the non-resonant states.\\
\\
We can summarize the discussion of the laser energy dependence of the $G$~peak intensity as follows:
In the very low energy regime, the intensity is suppressed due to angular momentum conservation
associated with the continuous rotation symmetry in the low-energy regime.
As the excitation energy increases, the intensity still remains low due to destructive quantum interference effects,
which leads to separate cancellations among both the resonant and non-resonant contributions to the Raman matrix element.
Finally, for larger laser energies ($\eL$$\gtrsim$2.5~eV), the resonant, in-phase contributions from the $K$-$M$ direction dominate the Raman matrix element.
Due to their increased weight as one approaches the van~Hove singularity at $M$, they cannot be suppressed by the sum of the non-resonant contributions.

\subsection{Fermi energy dependence}
\label{ssec:eF}

\begin{figure*}[htb]
    \includegraphics{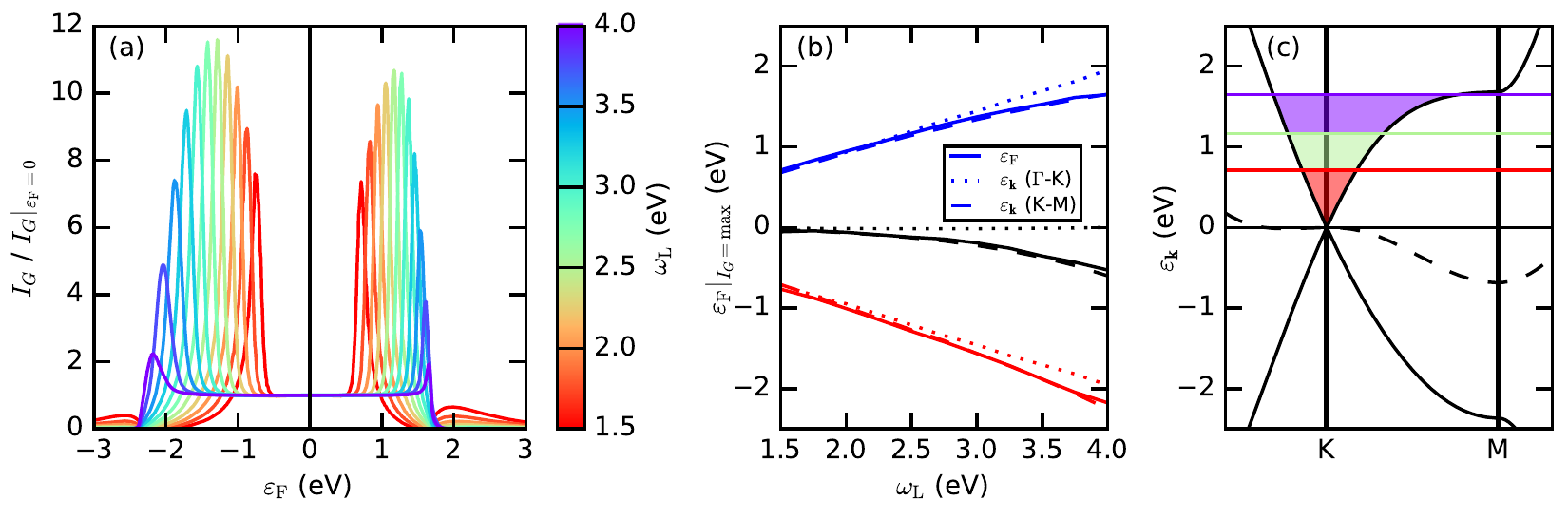}
    \caption{
        (a) $G$~peak intensity $I_G$ as a function of Fermi energy $\eF$ and laser energy $\eL$ (color-encoded),
            normalized to the intensity at zero Fermi energy.
        (b) Value of $\eF$ at which $I_G$ reaches a maximum for both electron (full blue line) and hole doping (full red line).
            The colored dashed (dotted) lines represent the respective conduction and valence band energy at the $\k$-point along the $K$-$M$
            ($\G$-$K$) direction that is in resonance with $(\oin+\oout)/2$.
            The black lines represent the sum of the respective blue and red lines, i.e., the electron-hole asymmetry.
        (c) $\pi$ and $\pi^*$~bands of graphene on part of the high-symmetry line $\G$-$K$-$M$-$\G$.
            The dashed line represents the electron-hole asymmetry of the band structure.
            The colored lines denote the positive Fermi levels which lead to the peak intensities as shown in panel (a)
            for $\eL$ = 1.5 (red), 2.5 (green), and 4~eV (violet).
            The shaded regions mark $\k$~points that do not contribute to the total Raman matrix element in the doped case,
            as the corresponding electronic transitions are blocked by the Pauli principle.
    }
    \label{fig:eF}
\end{figure*}

Next, we study the Fermi energy dependence of the $G$~peak intensity.
We varied the Fermi level $\eF$ from -3 to +3~eV relative to that of pristine graphene and calculated the resulting $G$~peak intensity
for different values of the laser energy $\eL$.
We use the rigid band approximation, i.e., we do not take into account the change of the electronic bands with $\eF$.
The results of the calculation, normalized to the respective intensity for pristine graphene, are shown in Fig.~\ref{fig:eF}a.
The different values of $\eL$ are represented by color, from $\eL$=1.5~eV (red) to $\eL$=4~eV (violet).
For both electron (right half) and hole doping (left half of the plot), the $G$~peak intensity shows a strong increase
when $\eF$ approaches a critical value.
This critical value corresponds well to the conduction or valence band energy of those states in the $K$-$M$~direction that are in resonance with
the average of the incoming and outgoing light energy, $(\oin$+$\oout)/2$, as shown in Fig.~\ref{fig:eF}b.
The strong behavior of $I_G$ as a function of $\eF$ has been observed in experiment~\cite{chen2011}, but so far only for one fixed laser energy.
As can be seen from Fig.~\ref{fig:eF}a, the strength of the relative increase of $I_G$ is expected to strongly vary with laser energy.
Furthermore, the electron-hole asymmetry (full black line in Fig.~\ref{fig:eF}b) of the $\eF$ values at which $I_G$ peaks
increases with excitation energy $\eL$, which can be understood from the electron-hole asymmetry of the band structure 
at the resonant $\k$-points in the $K$-$M$~direction (dashed black line in Fig.~\ref{fig:eF}b).

A conceptual explanation for the observed sensible dependence of $I_G$ on $\eF$ has already been suggested
in the original paper by Chen {\it et al.}~\cite{chen2011}.
The authors attributed the strong increase of $I_G$ at the critical values of $\eF$ 
to blocking of destructive quantum interference due to the Pauli principle.
When the Fermi level is increased (lowered), increasingly more transitions from the $\pi$ to the $\pi^*$~band are blocked,
as the corresponding electronic states in the conduction (valence) band become occupied (unoccupied).
As a result, these transitions become blocked by the Pauli principle.
In Fig.~\ref{fig:eF}c, this concept is illustrated for the case of electron doping and for three different laser energies.
The shaded regions mark those $\k$~points that do not contribute to the total Raman matrix element
as the corresponding electronic transitions are Pauli blocked.
In particular, if the Fermi level is tuned to certain values, a big part of destructive quantum interference effects is switched off
and as a result the $G$~peak intensity increases.
The strength of this increase, however, strongly depends on the laser energy.
To understand this last point, we consider the two extreme values of $\eL$ considered in our calculation, $\eL$=1.5 and 4~eV, in more detail.

For the case of $\eL$=1.5~eV, we see from Fig.~\ref{fig:eF}a that, on either side of charge neutrality point,
the intensity of the $G$~peak first goes through a strong increase, followed by a minimum,
and finally a smaller second peak that trails out as $\eF$ is increased even more.
This sequence of peaks and dips is easily understandable in terms of the picture of quantum interference
discussed for the ``mid-level'' energy regime in the previous section.
There we demonstrated the fact that in this $\eL$-regime, the total Raman matrix element is suppressed
because, on the one hand, almost-resonant contributions of states from $\k$-points around the resonance points mostly cancel each other,
while, on the other hand, non-resonant contributions from states around the van~Hove singularity are mostly canceled by
the sum of non-resonant contributions from the bulk of the BZ (see also the left-most panels of Fig.~\ref{fig:Mk} for illustration).
When the Fermi level is now increased from the charge neutrality point at $K$, more and more $\k$-points around $K$
do not contribute anymore to the total matrix element because of the Pauli principle.
In terms of the ``resonance peak'' picture in the $\Mk$ vs. $\k$-plot from Fig.~\ref{fig:Mk}, this means that at first,
an increase in $\eF$ blocks contributions from the ``inner'' flanks of the two resonance peaks,
which destructively interfere with the ``outer'' flanks.
This results in the observed strong increase of $I_G$.
The increase of $I_G$ continues until the Fermi level reaches the point where it blocks the resonant transitions,
at which point $I_G$ reaches a peak value (at approximately $\eF$=0.71~eV for electron doping).
An even further increase of $\eF$ results in a blocking of contributions from the constructively interfering ``outer'' flanks
and thus to a decrease of $I_G$ until a minimum intensity is reached (compare dip in red curve of Fig.~\ref{fig:eF}a at around $\eF$=1.55~eV),
corresponding to all almost-resonant transitions being Pauli blocked.
At this value of $\eF$, the only contributions to the total Raman matrix element come from the non-resonant states
from around the van~Hove singularity and from the bulk of the BZ, which cancel each other.
A further increase of $\eF$, however, leads to Pauli blocking of the van~Hove singularity contribution,
which can now no longer destructively interfere with the non-resonant contributions from the bulk of the BZ.
This leads to another increase of $I_G$ again, which peaks for a second time when $\eF$ reaches the van~Hove singularity at $M$
(see small bump in red curve of Fig.~\ref{fig:eF}a at around 1.97~eV).
An even further increase of $\eF$ only leads to more and more parts of the bulk of the BZ being Pauli blocked
and consequently $I_G$ trails out with increasing $\eF$.

By contrast, in the case of $\eL$=4~eV the $G$~peak intensity shows a much simpler and less strong behavior as a function of $\eF$.
When $\eF$ is increased away from the charge neutrality point, $I_G$ shows almost no response at first
before lightly peaking at a value of approx. $\eF$=1.65~eV.
After that, $I_G$ quickly reduces to zero when $\eF$ is increased even more.
This is again consistent with our earlier observation that for higher laser energies, the $G$~peak is mostly carried by resonant transitions
and quantum interference effects only play a minor role.
When $\eF$ is first increased, $I_G$ remains largely insensitive as only non-resonant transitions from around $K$ are Pauli blocked.
Only when $\eF$ approaches the resonant states near the van~Hove singularity at $M$ does $I_G$ feature a response,
when the destructive quantum interference effects from the ``inner'' flank (see right-most panel in Fig.~\ref{fig:Mk}) are switched off.
However, since the ``resonance peak'' in $\k$-space is broad and mostly in-phase,
the overall response of the $G$~peak intensity is not as strong for $\eL$=4~eV as for lower laser energy,
where the resonance peaks in $\k$~space are very narrow and quantum interference between the opposing flanks plays a much bigger role.
After $I_G$ for $\eL$=4~eV reached a peak value at around $\eF$=1.65~eV,
any further increase of the Fermi level only leads to the blocking of more and more resonant states,
leading to a sharp decrease of $I_G$ (compare violet curve in Fig.~\ref{fig:eF}a).
After the entire broad ``resonance peak'' in $\k$-space has been Pauli blocked,
the relative value of $I_G$ compared to the zero doping case remains insensitive to any further change of $\eF$,
confirming our previous finding that non-resonant states from the bulk of the BZ only play a minor role at higher laser energies.\\
\\
To summarize the Fermi energy dependence of the $G$~peak intensity,
our {\it ab initio} calculations confirm the suggestion of Chen {\it et al.} that the blocking of quantum interference effects
by shifting the Fermi level is the driving mechanism behind the strong increase of $I_G$ as $\eF$ reaches a critical value.
Going beyond this, we demonstrated that the relative increase of $I_G$ at these critical Fermi levels strongly depends on $\eL$,
as destructive quantum interference effects play an increasingly less dominant role with increasing $\eL$.
We also predict that for small laser energies,
the $G$~peak intensity shows a small resurgence even after all resonant states have already been blocked,
when $\eF$ approaches the van~Hove singularity,
thus blocking its destructive influence on the non-resonant contributions from the bulk of the BZ.

\subsection{Relevant states for the $G$~peak process}
\label{ssec:relstates}

Finally, we address a question first discussed for the low-energy case by Basko~\cite{basko2009},
namely which electronic transitions are relevant for the $G$~peak, i.e.,
which electronic states need to be taken into account in a theoretical description.
In the preceding two sections we demonstrated that the influence of quantum interference effects and of non-resonant states
strongly depends on the laser energy.
It is thus not surprising that the states that need to be considered for a quantitative description of the $G$~peak also vary with $\eL$.

\begin{figure}[h!]
    \includegraphics{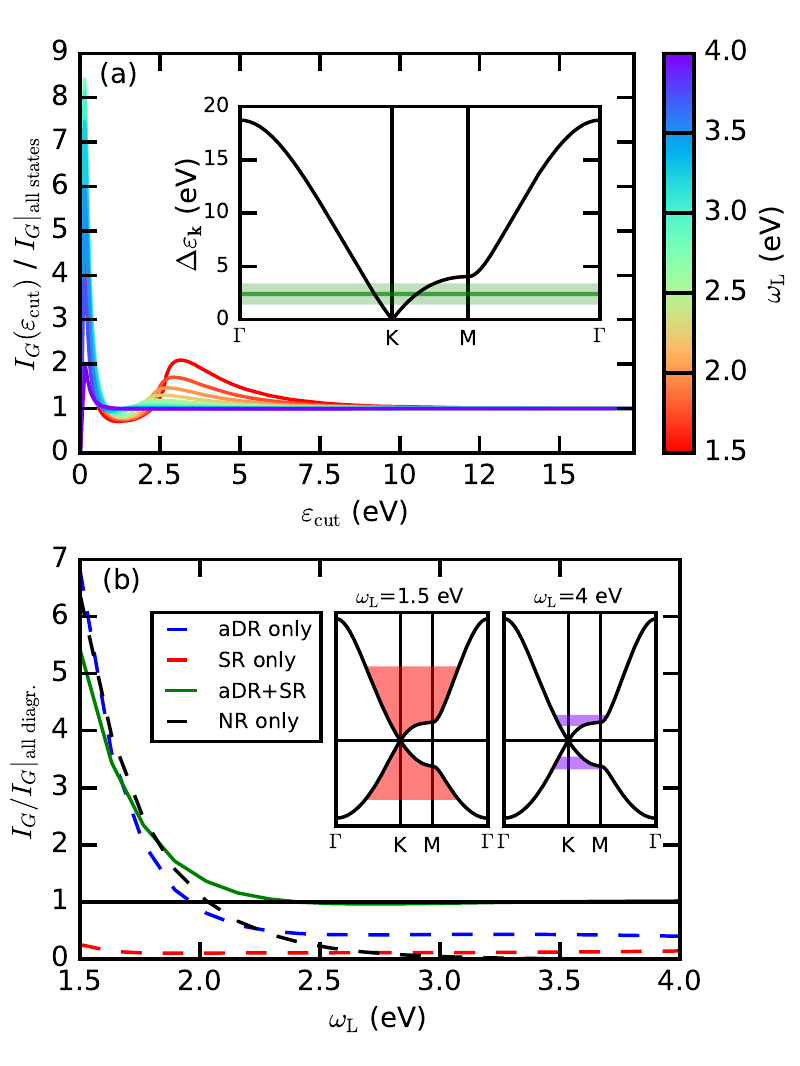}
    \caption{
        (a) $G$~peak intensity $I_G$ as a function of transition energy window width $\ecut$ as discussed in the text,
            normalized to the intensity calculated with all states included.
            The color corresponds to the laser energy $\eL$.
            Inset: Illustration of the transition energy window width.
            Dark shaded region: Window of states that are in resonance with the incoming or outgoing light frequency
            for a laser energy of $\eL$=2.5~eV. The width of the window is given by the electronic broadening $2\bargam$.
            Light shaded region: Included transition energy window for $\ecut$=1~eV.
        (b) $G$~peak intensity $I_G$ as a function of laser energy $\eL$ with certain resonant or non-resonant contributions only,
            normalized to the value of $I_G$ calculated with all contributions.
            The dashed blue, red, and black dashed lines correspond to $I_G$ calculated only with the aDR, SR, or NR contributions, respectively,
            The green line shows the values of $I_G$ calculated with the sum of the resonant aDR and SR terms,
            i.e., with $\M = \M^{\mr{aDR}} + \M^{\mr{SR}}$.
            Inset: Illustration of the relevant states (as defined in the text) for $\eL$ = 1.5 (left panel) and 4~eV (right panel).
    }
    \label{fig:relstates}
\end{figure}

To obtain a clear picture of which states contribute to the $G$~peak, we calculated the $G$~peak intensity for different $\eL$
with increasingly more states, starting with the ones that are in resonance with the incoming and/or outgoing light.
To be more precise, we introduce a transition energy window width $\ecut$ and include only those electronic transitions that obey the condition
\begin{equation}
\left| \Deleps - \frac{\oin+\oout}{2} \right| < \ecut,
\end{equation}
i.e., the included states lie within a window of width $2\ecut$ around the average resonance frequency $(\oin$+$\oout)/2$,
as illustrated in the inset of Fig.~\ref{fig:relstates}a.
We then proceed to calculate $I_G$ with only the states inside the selected window.
The resulting function $I_G(\ecut)$ is plotted for different laser energies $\eL$ in Fig.~\ref{fig:relstates}a.
For each $\eL$, we normalized with respect to the result obtained with all states included,
so that the all curves tend to one for large $\ecut$.

In Fig.~\ref{fig:relstates}a, we observe the same behavior as discussed in the previous two sections.
When $\ecut$ is first increased from zero, the first states to be included are the states
that are in resonance with the ingoing and/or outgoing light, which contribute with a large amplitude to the Raman matrix element.
As a result, $I_G(\ecut)$ displays a strong increase when the resonant states are first included.
The next states to be included stem from the flanks of the ``resonance peaks'' in $\k$-space (compare once more Fig.~\ref{fig:Mk}).
These states differ in phase from the resonant states and thus partially destructively interfere,
which leads to a decrease of $I_G(\ecut)$ when these states are included.

For low values of the laser energy $\eL$ (red to green curves in Fig.~\ref{fig:relstates}a), $I_G(\ecut)$
increases once more when $\ecut$ is increased.
This second peak corresponds to the inclusion of the states from around the van~Hove singularity,
which are all in-phase and thus constructively interfering.
Indeed, the point of steepest ascend in the red to green curves of Fig.~\ref{fig:relstates}a corresponds precisely to those values of $\ecut$
at which the energy window encompasses the $M$~point.
When $\ecut$ is increased even more, the destructively interfering states from the bulk of the BZ are taken into account
and hence $I_G(\ecut)$ decreases again.
When enough of these non-resonant states are taken into account,
$I_G(\ecut)$ finally converges to the value of $I_G$ calculated with all states.

For higher values of the laser energy though, $I_G(\ecut)$ does not change any more after the resonant states have been taken into account,
as seen in the behavior of the blue to violet lines in Fig.~\ref{fig:relstates}a, which converge straight to the correct value of $I_G$.
This once more shows that destructive interference from non-resonant states does not play a significant role in this $\eL$-regime.

To further prove this last point, we can consider the resonant and non-resonant terms separately
and calculate the $G$~peak intensity only with some of them.
Recall from Eq.~\ref{eq:Mres} that the total matrix element can be written as sum of an almost double-resonant (aDR),
a single-resonant (SR), and a non-resonant contribution (NR), with each one of them already implicitly summed over $\k$-points.
If we calculate $I_G$ with only the aDR-, the SR-, or the NR-contribution as a function of $\eL$ and normalize to the value of $I_G$
including all contributions, we obtain the dashed lines shown in Fig.~\ref{fig:relstates}b.
Doing the calculation only with the aDR- or the NR-terms severely overestimates the $G$~peak intensity for low excitation energies,
while it underestimates the correct result for $I_G$ for larger $\eL$.
A calculation with only the single-resonant terms, on the other hand, always underestimates the correct value for $I_G$.

More interesting, however, is the result we obtain when we consider the sum of the aDR- and SR-contributions in the total matrix element,
i.e., we include all terms in the total matrix element that can go resonant with either the incoming or outgoing light.
The resulting curve is shown as a green line in Fig.~\ref{fig:relstates}b.
The plot confirms our previous analysis in that, for higher values of the laser energy ($\eL$$\gtrsim$2.5~eV),
the resonant contributions are the dominating ones.
Indeed, a calculation that only takes into account the resonant terms in the Raman matrix element yields very good results in this laser energy regime.

As a last point, we can finally visualize the required states for different values of $\eL$
by reconsidering our result obtained with a transition energy window width $\ecut$ shown in Fig.~\ref{fig:relstates}a.
If we start from the maximum considered value of the cutoff (the full $\pi$-band width), and gradually lower the value of $\ecut$
until $I_G(\ecut)$ differs from the correct value of $I_G$ by more than 2\%, we obtain a minimum transition energy window width
around the resonant states that are needed to achieve 2\% accuracy.
For this minimal value of $\ecut$, we can visualize the corresponding $\k$-points within this transition energy window
in a plot of the band structure.
This is shown for the two laser energies $\eL$=1.5 and $\eL$=4~eV in the inset of Fig.~\ref{fig:relstates}b.
In this representation it becomes immediately clear, that for low laser energies
the relevant electronic transitions for the $G$~peak come from a large part of the BZ,
in agreement with the findings of Basko~\cite{basko2009} in his analytical work in the low-energy limit.
For larger values of $\eL$, this picture changes, however,
in that the relevant states are localized in a broad band of total width $\sim$2~eV around the resonance energy.
It is important to note that in neither $\eL$-regime is it sensible to approximate the Raman amplitude
with an expression that only takes into account resonant electronic transitions
(in an energy window of $\pm$ the inverse electronic life time).


\section{Conclusions}

In conclusion, we developed a fully {\it ab initio} approach to the calculation and study of one-phonon Raman intensities
and applied it to the case of monolayer graphene.
We demonstrated explicitly that quantum interference effects play a crucial role for the understanding and interpretation
of the Raman process leading to the $G$~peak of graphene.
In particular, quantum interference effects from non-resonant states are very important
for a correct low-excitation energy description of the $G$~peak.
We confirmed the conceptual picture of Chen {\it et al.}~\cite{chen2011} that Pauli blocking of destructive quantum interference
by tuning the Fermi level to a critical value can lead to a large enhancement of the $G$~peak intensity.
Furthermore, we predict that the relative increase of the $G$~peak intensity at the critical Fermi level
strongly depends on the laser energy.
Finally, our general approach enabled us to explicitly demonstrate which electronic states need to be taken into account
for an accurate description of the $G$~peak for a wide range of laser energies,
 going beyond the previous analytical analysis by Basko~\cite{basko2009} in the low-energy regime.
In particular, we showed that for low laser energy, transitions from almost the entire first Brillouin zone need to be considered,
whereas at higher laser energy, only states in a broad band of full width $\sim$2~eV around the resonance energy need to be taken into account.

While in this work we focused on a study of one-phonon Raman intensities in monolayer graphene,
our implementation and method of analysis is completely general.
As such, it can be applied to any material and enables the study of other materials as well.


\section{Acknowledgments}

The authors would like to thank H.P.C.~Miranda for helpful discussions.
S.R. and L.W. acknowledge financial support by the National Research Fund (FNR) Luxembourg (projects RAMGRASEA and TMD-nano).



%

\end{document}